\documentclass[fleqn]{article}
\usepackage{graphicx}

\begin{document}
\title{
 Pad\'e-Improved Estimates  of Hadronic Higgs Decay Rates}

\author{
V. Elias and F. A. Chishtie\\
{\sl Department of Applied Mathematics}\\
{\sl The University of Western Ontario}\\
{\sl London, Ontario  N6A 5B7  CANADA}
\\[10pt]
T. G. Steele\\
{\sl Department of Physics and Engineering Physics}\\
{\sl University of Saskatchewan}\\
{\sl Saskatoon, Saskatchewan  S7N 5E2   CANADA}
}

\maketitle

\begin{abstract}Asymptotic Pad\'e-approximant methods are utilized to estimate the 
${\cal{O}}(\alpha_s^5)$ contribution to the $H \rightarrow gg$ rate and the 
${\cal{O}}(\alpha_s^4)$ contribution to the $H \rightarrow b \bar{b}$
rate.  The former process is of particular interest because of the slow
convergence evident from the three known terms of its QCD series, which
begins with an ${\cal{O}}(\alpha_s^2)$ leading-order term.  The 
${\cal{O}}(\alpha_s^5)$ contribution to the $H \rightarrow gg$ rate is
expressed as a degree-3 polynomial in $L \equiv ln (\mu^2 / m_t^2
(\mu))$.  We find that asymptotic Pad\'e-approximant predictions for the
coefficients of $L$, $L^2$, and $L^3$ are respectively within 1\%, 2\%,
and 7\% of true values extracted via renormalization-group methods.
Upon including the full set of next-order coefficients, the $H
\rightarrow gg$ rate is found to be virtually scale-independent over the
0.3 $M_{H} \stackrel{<}{_\sim} \mu \stackrel{<}{_\sim} M_t$ range of
the renormalization scale-parameter $\mu$.  We conclude by discussing
the small ${\cal{O}}(\alpha_s^4)$ contribution to the $H \rightarrow b
\bar{b}$ rate, which is obtained from a prior asymptotic 
Pad\'e-approximant estimate of the ${\cal{O}}(\alpha_s^4)$ contribution 
to the quark-antiquark scalar-current correlation function.
\end{abstract}

\section*{1.  Introduction} 

The Higgs boson characterizes the electroweak symmetry breaking
underlying the Standard Model.  Its discovery and phenomenology will be
of immense importance in clarifying our understanding of this symmetry
breaking, as well as in providing vital information as to the nature of
beyond-the-Standard-Model physics.  The two leading hadronic decay modes
of a Weinberg-Salam Higgs boson $(H)$ with mass between $100 \; {\rm GeV}$ 
and $160 \; {\rm GeV}$ are the QCD processes $H \rightarrow b
\overline{b}$ and $H \rightarrow$ two gluons $(gg)$.  Although the rate
for this latter process is known to ${\cal{O}}(\alpha_s^4)$, such precision
incorporates only two non-leading orders of a slowly converging series
in the strong coupling.  If $M_H = 100 \; {\rm GeV}$, for example, the
known order-by-order QCD corrections have been calculated \cite{kgc} to
be $1 + 0.66 + 0.21$.

The work presented here is primarily directed toward obtaining a more
precise estimate of the $H \rightarrow gg$ decay rate. We utilize
renormalization-group $(RG)$ and asymptotic Pad\'e-approximant methods
to estimate the full next-order contribution to the underlying
correlation function for this process.  Such an approach
has already been applied to the quark-antiquark scalar-current
correlation function underlying the $H \rightarrow b \overline{b}$ rate
\cite{ces}, a calculation we review in Section 5 of the present paper. 

As in our prior analysis of the two-gluon decay amplitude of a non-Standard-Model 
CP-odd Higgs field \cite{fac},
the approach we take here is to test asymptotic Pad\'e-approximant
estimates against RG-accessible coefficients within the next-order of
perturbation theory.  We show such estimates to be accurate up to relative
errors of only a few percent, supporting the credibility of the
same approach in estimating the 
RG-{\it inaccessible} coefficient needed to determine the full
${\cal{O}}(\alpha_s^5)$ contribution to the $H \rightarrow gg$ rate.

We operate within the context of $\overline{MS}$ expressions for the $H \rightarrow gg$
and $H \rightarrow b \bar{b}$ rates that explicitly depend on an arbitrary
renormalization scale $\mu$.  It has been argued
elsewhere \cite{egk,ega,mra} that asymptotic Pad\'e-approximant methods reduce
the explicit scale-dependence of perturbative quantities which must ultimately be scale
invariant.  We find this to be the case for the $H \rightarrow gg$ rate
as well, despite residual scale-sensitivity anticipated from the estimated 
RG-inaccessible coefficient of the nonlogarithmic ${\cal{O}} (\alpha_s^5)$ term 
within the next-order correlator.

In Section 2, we demonstrate the explicit RG-invariance of the $H \rightarrow gg$
rate, as calculated in \cite{kgc} in the $m_b \rightarrow 0, \; M_H^2 < < 4 M_t^2$ 
limit. This RG invariance enables one to calculate all the next-order 
coefficients $c_k$ of  $ln^k [\mu^2/m_t(\mu^2)]$  $(\alpha_s (\mu) / \pi)^5$
within the calculated rate; only the $k=0$ constant term is RG-inaccessible.

In Section 3, we review how asymptotic Pad\'e-approximant methods may be
utilized to estimate this set of next-order coefficients, and demonstrate
close agreement with the RG-determinations of Section 2 over the range 
$100 \leq M_H \leq 175 \; {\rm GeV}$.  The RG-inaccessible coefficient $c_0$ is 
also estimated over this range of Higgs masses and is fitted to its anticipated
behaviour as a degree-3 polynomial in $ln [(M_H^2 / M_t^2)_{pole}]$.

In Section 4 we examine the residual scale dependence of the 
Pad\'e-improved $H \rightarrow gg$ decay rate, as obtained in Section 3.
Although we anticipate a relative scale dependence 
comparable to $c_0(\alpha_s (\mu) / \pi)^3 (\approx 3
- 10 \%)$, since $c_0$ cannot be extracted perturbatively from the RG
equation (to the order we consider), we find that the residual scale
dependence of the decay rate is a full order of magnitude smaller 
than this estimate for $\mu$ between 0.3 $M_H$ and $M_t$ (we assume
$100 \; {\rm GeV} \leq M_H \leq 175 \; {\rm GeV}$).
Consequently, the scale-dependence of the
Pad\'e-derived term $c_0(\alpha_s (\mu) / \pi)^5$ very nearly cancels
the scale-dependence anticipated from truncation of the
perturbative series for the $H \rightarrow gg$ rate, thereby facilitating the
(nearly) scale-invariant rate predictions of Table 6.  Predictions of
the $H \rightarrow gg$ decay rate are tabulated both for the $m_b
\rightarrow 0, \; M_H^2 < < 4M_t^2$ limit, and for the leading-order
departures from this limit for appropriate constant and running values for $m_b$ and
$M_t$.

In the concluding section, we review how prior asymptotic 
Pad\'e-approximant estimates of the ${\cal{O}}(\alpha_s^4)$ contribution to the
scalar-current correlation function \cite{ces} can be incorporated into
the $H \rightarrow b \overline{b}$ decay rate.  We demonstrate that the
resulting next-order contribution, though only 0.01\% of the leading term, 
is somewhat {\it larger} than known ${\cal{O}}(\alpha_s^2)$ power-suppressed
contributions \cite{che}.  Consequently, the estimated ${\cal{O}}(\alpha_s^4)$
term enables the $H \rightarrow b \bar{b}$ rate to be estimated to
four significant figure accuracy.

\section*{2.  RG-Invariance of the $H \rightarrow gg$ Rate}

\renewcommand{\theequation}{2.1\alph{equation}}
\setcounter{equation}{0}

The Higgs $\rightarrow$ gg decay rate is given explicitly 
to 3-loop order in \cite{kgc} by the following expressions:
\begin{equation}
\Gamma_{H \rightarrow gg} =\frac{\sqrt{2}G_F}{M_H} 
R(\alpha_s, q^2 = M_H^2, \mu^2, M_t^2),
\end{equation}

\begin{equation}
R=C_1^2 Im <[0_1^\prime ]^2>,
\end{equation}

\begin{eqnarray}
C_1=-\frac{x^{(6)}}{12}\biggl[ 
1&+&x^{(6)}\left( \frac{11}{4}-\frac{1}{6}
ln\left(\frac{\mu^2}{M_t^2}\right)\right)
\biggr.\nonumber \\
\biggl.& +&(x^{(6)})^2 \left(\frac{211}{36}+\frac{55}{48} ln
\left(\frac{\mu^2}{M_t^2}\right)
+  \frac{1}{36} ln^2
\left(\frac{\mu^2}{M_t^2}\right) \right)+{\cal{O}}\left[(x^{(6)})^3\right]\biggr]
\end{eqnarray}

\begin{eqnarray}
Im<[0_1^\prime ]^2>
= \frac{2q^4}{\pi} \biggl[ 1 &+& x^{(5)}\left( \frac{149}{12}+\frac{23}{6}
ln \left( \frac{\mu^2}{q^2} \right) \right) \biggr.\nonumber \\
&+&(x^{(5)})^2 \left( 68.64817+\frac{1297}{16} ln
\left(\frac{\mu^2}{q^2}\right)+\frac{529}{48}ln^2 \left(
\frac{\mu^2}{q^2}\right) \right) 
+  {\cal{O}}\left[(x^{(5)})^3\right] \biggr].
\end{eqnarray}
In the above expression, $<[0_1^\prime ]^2>$ is the vacuum polarization
of the Higgs field induced via the gluon operator $0_1^\prime = G_{a\mu\nu} G_a^{\mu\nu}$
\cite{kgc}, and
$x^{(n_f)}\equiv \alpha_s^{(n_f)}/\pi$, where
$\alpha_s^{(n_f)}(\mu)$ is the running strong coupling with $n_f$ active
flavours.  Five flavours are assumed to be light in both (2.1c) and
(2.1d).  The $t$-quark mass $M_t$ appearing in (2.1c) is an RG-invariant pole
mass, and $M_H^2$ is assumed to be small compared to $4M_t^2$ \cite{kgc}.

Our goal here is to use asymptotic Pad\'e-approximant methods in
conjunction with the RG-invariance of (2.1a) in order to predict the
next order contribution to $\Gamma(H \rightarrow gg)$.  In a previous
application \cite{mra} of such methods to the inclusive semileptonic $b
\rightarrow u$ rate, it was found that success in predicting RG-accessible
next-order coefficients was greatly enhanced by recasting the
entire expression in terms of the running fermion mass.  Indeed, such
replacement of the $b$-quark pole mass with its scale-dependent $\overline{MS}$ 
mass had already been employed by van Ritbergen \cite{tur} to avoid a
renormalon pole.

Consequently, we first re-express the $H \rightarrow 2g$
rate in terms of the running $t$-quark mass $m_t(\mu)$,
which evolves via a six-active-flavour $\gamma_m$-function, and the
corresponding six-active-flavour running coupling $x^{(6)}(\mu)$.  This
transformation is facilitated by the following relationships \cite{sal}:

\renewcommand{\theequation}{2.\arabic{equation}}
\setcounter{equation}{1}

\begin{equation}
x^{(5)}(\mu)=x^{(6)}(\mu)-\frac{(x^{(6)}(\mu))^2}{6} ln
\left(\frac{\mu^2}{m_t^2 (\mu)} \right) +
{\cal{O}}\left[(x^{(6)})^3\right],
\end{equation}

\begin{equation}
m_t(\mu)/M_t=\left[ 1-x^{(6)}(\mu) \left( \frac{4}{3} + ln \left(
\frac{\mu^2}{M_t^2}\right) \right)+{\cal{O}}\left[(x^{(6)})^2 \right] \right. .
\end{equation}
Equation (2.3) implies that the logarithm in (2.1c) may be re-expressed
in terms of $L \equiv ln (\mu^2/m_t^2 (\mu))$:

\begin{equation}
ln \left( \frac{\mu^2}{M_t^2} \right) = L - x^{(6)}(\mu)\left[ 2L +
8/3\right] + {\cal{O}}\left[ (x^{(6)})^2\right]
\end{equation}
With $q^2=M_H^2$, the logarithm in (2.1d) can also be
expressed entirely in terms of the running $t$-quark mass and a
logarithm $T\equiv ln \left( M_H^2 / M_t^2\right)$ of the ratio of 
$RG$-invariant pole masses:
\begin{equation}
ln \left( \frac{\mu^2}{M_H^2} \right) = L-T - x^{(6)}(\mu) \left[
2L+8/3\right] + {\cal{O}} \left[ (x^{(6)})^2\right].
\end{equation}
Substitution of (2.2) and (2.5) into (2.1d) with $q^2=M_H^2$, and
substitution of (2.4) into (2.1c) leads to the following expression for
the $H \rightarrow gg$ decay rate:
\begin{equation}
\Gamma_{H\rightarrow gg} = \frac{\sqrt{2} G_F M_H^3}{72\pi} S \left[
x^{(6)}(\mu), L(\mu), T \right],
\end{equation}
\begin{eqnarray}
S\left[x,L,T\right]=x^2 \biggl( 
1 &+& x \left[ \left( \frac{215}{12} -
\frac{23T}{6} \right) + \frac{7}{2} L \right] 
\biggr. \nonumber \\
& +& x^2 \left[ \left( 146.8912 - \frac{4903}{48} T + \frac{529}{48} T^2
\right) 
+  \left( \frac{1445}{16} - \frac{161}{8} T \right) L + \frac{147}{16}
L^2 \right] 
\nonumber \\
 \biggl. &+&  x^3 \left[ c_0 + c_1 L + c_2 L^2 + c_3 L^3 \right] + {\cal{O}}(x^4)\biggr).
\end{eqnarray}

In (2.7), we list the unknown coefficients $c_0, c_1, c_2$ and $c_3$ of
the 4-loop contribution to the rate.  Three of these may be extracted by
the scale-[RG-] invariance of the physical decay rate:  $d\Gamma/d\mu =
0$.  This invariance implies that 
\begin{eqnarray}
O = \mu \frac{dS}{d\mu} [x,L,T], \nonumber \\
= [1 - 2\gamma_m (x)] \frac{\partial S}{\partial L} + \beta(x)
\frac{\partial S}{\partial x} .
\end{eqnarray}
Both the $\beta$ and $\gamma_m$ functions in (2.8) are referenced to
six active flavours:
\begin{equation}
\beta^{(6)}(x) = -\frac{7}{4} x^2 - \frac{13}{8} x^3 + \frac{65}{128}
x^4 ...
\end{equation}
\begin{equation}
\gamma_m^{(6)} (x) = -x - \frac{27}{8} x^2 ...\; \; \;  .
\end{equation}
One can easily verify from the known terms listed in (2.7) that (2.8) is
perturbatively valid to orders $x^3, x^4$ (including explicit
cancellation of terms involving $T$), and $x^4 L$.  The continued
perturbative validity of (2.8) to orders $x^5 L^2$, $x^5 L$, and $x^5$
is sufficient to determine the four-loop coefficients $c_1, c_2$, and
$c_3$:
\begin{equation}
c_1 = 910.3167 - \frac{16643}{24} T + \frac{3703}{48} T^2,
\end{equation}
\begin{equation}
c_2 = \frac{1225}{4} - \frac{1127}{16} T,
\end{equation}
\begin{equation}
c_3 = \frac{343}{16}.
\end{equation}
The coefficient $c_0$ is not RG-accessible to these orders.  In the next
section, we will utilize asymptotic Pad\'e approximant methods to
estimate the four-loop coefficients $\{c_0, c_1, c_2, c_3 \}$.  As
in prior work \cite{ces,fac,mra}, the accuracy of these predictions in
reproducing (2.11 - 2.13) will serve as an indication of the accuracy of
our estimate for $c_0$.

\section*{3. Pad\'e-Predictions for the Four-Loop\\ \indent Coefficients}

\renewcommand{\theequation}{3.\arabic{equation}}
\setcounter{equation}{0}

The asymptotic Pad\'e-approximant procedure for estimation of the four loop
coefficients $c_0, c_1, c_2, c_3$ in (2.7) has been delineated in
previous work \cite{ces,fac,mra}.  The series (2.7) may be expressed in
the form

\begin{equation}
S[x,L,T] = x^2 \left[ 1 + R_1 [L,T] x + R_2 [L,T]x^2 + R_3[L,T]x^3 + ...
\right]
\end{equation}
where $R_1$ and $R_2$ are known, and $R_3$ is to be determined:
\begin{equation}
R_1[L,T] = \left( \frac{215}{12} - \frac{23}{6} T \right) + \frac{7}{2}L
\end{equation}

\begin{equation}
R_2 [L,T] = 146.8912 - \frac{4903}{48}T + \frac{529}{48}T^2 
+ \left( \frac{1445}{16} - \frac{161}{8} T \right)L + \frac{147}{16} L^2
\end{equation}

\begin{equation}
R_3 [L,T] = c_0(T) + c_1(T)L + c_2(T)L^2 + c_3 L^3 .
\end{equation}
Initially, we shall eliminate $T$ as a variable by assuming that the
Higgs pole mass is 100 GeV, in which case $T = ln [M_H^2 / M_t^2] = 2 \; ln
(100/175.6) = -1.126$.  As described in \cite{mra} and \cite{ejj}, the
$[0|1]$ Pad\'e-approximant prediction for $R_2$ is
\begin{equation}
R_2^{[0|1]} = R_1^2
\end{equation}
and the $[1|1]$ Pad\'e-approximant prediction for $R_3$ is
\begin{equation}
R_3^{[1|1]} = R_2^2/R_1 .
\end{equation}
If the error of $[N|1]$ approximants in predicting $R_{N+2}$, the $N+2$
term in the perturbative series, is inversely proportional to $N+1$ \cite{ejj}
--- {\it i.e.}, if 
\begin{equation}
\frac{R_{N+2}^{[N|1]} - R_{N+2}^{exact}}{R_{N+2}^{exact}} = \frac{-
A}{N+1},
\end{equation}
where $A$ is a constant--- one may then utilize (3.5) and the exact value for
$R_2$ within (3.7) to obtain $A = (R_2^2 - R_1^2)/R_2$.  Substituting (3.6) and 
this estimate for $A$ into (3.7), one obtains an error-improved estimate
for the unknown coefficient $R_3$ \cite{esc}:
\begin{equation}
R_3[L] = 2R_2^3 [L] / (R_1[L] R_2[L] + R_1^3 [L])
\end{equation}
To obtain estimates of the coefficients $c_0,c_1,c_2,c_3$ within (3.4), we match
the scale dependence of (3.4) to that of (3.8) over the purely
perturbative $L > 0$ region [corresponding to the ultraviolet scales
$\mu > m_t(\mu)$] through use of the moment integrals
\begin{equation}
N_k \equiv (k+2) \int_0^1 dw \; w^{k+1} R_3 (w),
\end{equation}
where $w = m_t^2 (\mu) / \mu^2\;  [L = -ln (w) ]$.  Substitution of (3.4)
into the integrand of (3.9) yields the following expressions for the
first four moments \cite{ces}:

\renewcommand{\theequation}{3.10\alph{equation}}
\setcounter{equation}{0}

\begin{equation}
N_{-1} = c_0 + c_1 + 2c_2 + 6c_3,
\end{equation}
\begin{equation}
N_0 = c_0 + \frac{1}{2}c_1 + \frac{1}{2}c_2 + \frac{3}{4}c_3,
\end{equation}
\begin{equation}
N_1 = c_0 + \frac{1}{3}c_1 + \frac{2}{9}c_2 + \frac{2}{9}c_3,
\end{equation}
\begin{equation}
N_2 = c_0 + \frac{1}{4}c_1 + \frac{1}{8}c_2 + \frac{3}{32}c_3.
\end{equation}
However, explicit numerical estimates of these four moments may be
obtained via substitution of (3.2) and (3.3) for the respective factors
of $R_1[L]$ and $R_2[L]$ appearing in (3.8), and by subsequent substitution
of this estimate for $R_3[L]$ into the integrand of (3.9) [with $L = -ln
(w)$].  For $M_H = 100 \; {\rm GeV}$ the resulting estimates are 

\renewcommand{\theequation}{3.1\arabic{equation}}
\setcounter{equation}{0}
\begin{equation}
N_{-1} = 5102.9, \; N_0 = 3542.9, \; N_1 = 3131.8,  \; N_2 = 2944.7.  
\end{equation}
Substitution of these values into (3.10) yields the following predicted values
$(c_i^{Pad\acute{e}})$ for the four-loop terms in the $H \rightarrow gg$ decay
rate (2.7):
\begin{equation}
c_0^{Pad\acute{e}} = 2453, \; c_1^{Pad\acute{e}} = 1772, \;
c_2^{Pad\acute{e}} = 378.4, \; c_3^{Pad\acute{e}} = 20.27.
\end{equation}
The exact values of $c_1, c_2$, and $c_3$ were determined via $RG$
methods in the previous section.  For $M_H = 100 \; {\rm GeV}$ [{\it i.e.} for $T
= -1.126$], these values are found from (2.11-13) to be 
\begin{equation}
c_1 = 1789.02, \; c_2 = 385.568, \;  c_3 = 21.4375.
\end{equation}
Comparing (3.12) and (3.13), one finds the relative error 
$\left[ \delta c_i \equiv (c_i^{Pad\acute{e}} - c_i) / c_i\right]$ 
of the Pad\'e estimates for $c_1, c_2$ and $c_3$ to be -0.95\%, -1.9\%, 
and -5.5\%, respectively.

Such accuracy in predicting the known four-loop terms in the $H
\rightarrow gg$ rate suggests that the prediction for $c_0$ in (3.12) is
a credible one.  One way to test the stability of this prediction is to
utilize the true values of $c_1, c_2, c_3$ within the moment expressions
(3.10) to obtain [for the numerical values (3.11) already obtained for
these moments] four independent determinations of $c_0$.  We then find
that
\begin{eqnarray}
(3.10a): \; \; c_0 = 2412, \nonumber \\
(3.10b): \; \; c_0 = 2436, \nonumber \\
(3.10c): \; \; c_0 = 2444, \nonumber \\
(3.10d): \; \; c_0 = 2447,
\end{eqnarray}
results all within 2\% of that in (3.12).

An alternative approach to matching the four-loop coefficients $c_i$
within (3.4) to the error improved Pad\'e estimate (3.8) is optimize
the least-squares function \cite{fac}
\begin{equation}
\chi^2 [c_0, c_1, c_2, c_3] = \int_0^1 \left[ R_3 - (c_0 - c_1 ln w +
c_2 ln^2 w - c_3 ln^3 w)\right]^2 dw,
\end{equation}
with $R_3$ in the integral given by (3.8).  As before, factors of $R_1$
and $R_2$ appearing in (3.8) are given explicitly by (3.2) and (3.3)
with $L = -ln (w)$.  One then finds for $M_H = 100 \; {\rm GeV}$ $[T = -1.126]$
that
\begin{eqnarray}
\chi^2(c_0,c_1,c_2,c_3)&=&4.064168878 \cdot 10^7 + 720 c_3^2 
+4 c_0 c_2 + c_0^2 + 2c_0 c_1 + 24 c_2^2 
+ 12 c_1 c_2 + 12 c_0 c_3 
+2c_1^2 \nonumber\\
&+& 240 c_2 c_3 
+ 48 c_1 c_3 - 10205.87 c_0 
-17507.05 c_1 - 54106.23 c_2 - 234536.67 c_3 
\end{eqnarray}
The optimization requirement
\begin{equation}
\frac{\partial \chi^2}{\partial c_i} = 0
\end{equation}
yields predictions remarkably close to those of (3.12),
\begin{equation}
c_0^{\chi^2} = 2452, \; \; c_1^{\chi^2} = 1774, \; \; c_2^{\chi^2} 
= 377.2, \; \; c_3^{\chi^2} = 20.45,
\end{equation}
further confirming the stability of the estimation procedure.

In Table 1, we have tabulated a set of predictions of the four-loop term
$c_0$ for values of the Higgs mass between 100 and 175 GeV.  Also
tabulated are the errors in the predicted values for $c_1, c_2, c_3$,
relative to the true values for these coefficients, as given in 
(2.11-13).  Estimated values of $c_1$ and $c_2$ remain within 2\% of their
true values (2.11-13) over the range of Higgs masses given;  the
relative error of $c_3$ estimates remains below 7\% for the same range.
This consistency provides further support for the $c_0$ estimates presented 
in the final column.

These $c_0$ estimates may be utilized to ascertain the convergence of
the series (2.7).  If we choose $\mu = m_t (\mu)$, the logarithmic
factors $L^k$ within (2.7) all vanish.  To evaluate the series $S/x^2$
in (2.7), we assume that $m_t(m_t) \cong M_t = 175.6 \; {\rm GeV}$ and that
$\alpha_s(175.6) = 0.10915 = \pi x(175.6)$, as evolved from $\alpha_s
(M_z) = 0.119$ \cite{cea}, in which case
\begin{equation}
S(\mu = 175.6 \,{\rm GeV}) / x^2 (175.6) 
=1 + a_0 x(175.6) + b_0 x^2 (175.6) 
+ c_0 x^3 (175.6).
\end{equation}
For Higgs masses between 100-175 GeV, we tabulate in Table 2 the
magnitudes of successive orders in (3.19).  The final 4$^{th}$ order term
is obtained from the appropriate Pad\'e estimate for $c_0$ listed in
Table 1.  Table 2 shows that the four-loop term decreases from 10\% to
3\% of the leading contribution as $M_H$ increases from 100
to 175 GeV.  Moreover, the convergence of the series is problematical in
the absence of the estimate for the four loop term.  Over the range of
Higgs masses considered, the three-loop contribution is between 18\% and
33\% of the leading (one-loop) contribution.

\renewcommand{\theequation}{3.2\arabic{equation}}
\setcounter{equation}{-1}

The procedures delineated above can also be utilized to predict $c_0$'s
explicit polynomial dependence of $T$,
\begin{equation}
c_0 (T) = a_0 + a_1 T + a_2 T^2 + a_3 T^3,
\end{equation}
analogous to the expressions (2.11-13) obtained from $RG$-invariance for
$c_1(T)$, $c_2(T)$, and $c_3(T)$.  To extract the $T$ dependence of
$c_0$, we first incorporate all known $T$-dependence into the least-
squares function (3.15):
\begin{eqnarray}
\chi^2 [c_0(T)] = \int_0^1 \left[ R_3[L,T] - \left\{ c_0(T)+c_1(T)L 
+  c_2 (T) L^2 + c_3 (T) L^3 \right\} \right]^2 dw.
\end{eqnarray}
In (3.21), $L \equiv -ln w$.  The quantities $c_1(T)$, $c_2(T)$ and $c_3(T)$ are no
longer optimizable variables as in (3.15), but are now the explicit
polynomials (2.11-13) obtained in the previous section via $RG$-methods.
The factor $R_3 [L,T]$ in the integrand of (3.21) is just (3.8)
generalized to include the explicit $T$-dependence of $R_1$ and
$R_2$ [eqs. (3.2) and (3.3)]:
\begin{equation}
R_3 [L,T] = \frac{2 R_2^3 [L,T]}{R_1 [L,T] R_2 [L,T] + R_1^3 [L,T]}.
\end{equation}
From (3.21), the requirement $d \chi^2 / d c_0 = 0$ generates $c_0$ as a
function of $T$.  Since $T \equiv ln (M_H^2 / M_T^2)$, we restrict our
attention to the region $-1 \leq T \leq 0$ ({\it i.e.} to values for $M_H$
between $M_t$ and $M_t e^{-1/2} = 107 \; {\rm GeV}$).  A set of values for
$c_0 (T)$ can be obtained via optimization of (3.21) over values of $T$ in
this region:
\begin{eqnarray}
c_0(0) & = & 735.7, \; c_0(-0.1) = 841.7, \; c_0(-0.2) = 955.6, \; c_0(-0.3) = 1078, \nonumber \\
c_0(-0.4) & = & 1208, \; c_0 (-0.5) = 1346, \; c_0(-0.6) = 1493, \;  c_0(-0.7)=1649, \nonumber \\
c_0(-0.8) & = & 1814, ;\  c_0(-0.9) = 1988, \; c_0(-1.0) = 2170.
\end{eqnarray}
We obtain a least-squares fit of these results to the form (3.20)  by
optimizing
\begin{equation}
\chi^2[a_0, a_1, a_2, a_3] \equiv \sum_{i=0}^{10} 
\left[ c_0 (-i/10) - \left( a_0 - \frac{a_1 i}{10} 
+ \frac{a_2 i^2}{100} - \frac{a_3 i^3}{1000} \right) \right]^2
\end{equation}
with respect to $\{a_0, a_1, a_2, a_3 \}$, and find that
\begin{equation}
c_0(T) = 735.7 - 1020 T + 388.8 T^2 - 25.41 T^3.
\end{equation}
Each coefficient listed above is within 5\% relative error of the
corresponding coefficient obtained via a least squares fit of the $c_0$
values displayed in Table 1 to the polynomial form (3.20):
\begin{equation}
c_0(T) = 755.9 - 1029 T + 394.3 T^2 - 26.74 T^3.
\end{equation}

Note that (2.11), (2.12), (2.13) and the
Pad\'e-estimates (3.25) or (3.26) specify {\it all} the logarithmic
coefficients within the full four loop contribution (3.4) to the $H \rightarrow
gg$ decay rate (2.6,7), and comparison of these results with future perturbative 
calculations should yield information which can be employed to further improve
Pad\'e estimation procedures. The exact $H \rightarrow gg$ rate (2.6) is 
necessarily a scale invariant physical quantity.
The factor $S [x(\mu), L(\mu), T]$ within (2.6) will exhibit residual
scale-dependence [{\it i.e.} $\mu$-dependence] only as a consequence of
truncation of the series $S$ to a given order of perturbation theory.
In the section that follows, the incorporation
of the four-loop coefficients $c_{0-3}$ of $S [x(\mu), L(\mu),T]$
is seen to eliminate virtually all of this residual scale dependence.

\section*{4. The $H \rightarrow gg$ Rate}

In Fig. 1, the $\mu$-dependence of three- and
four-loop expressions for $S[x^{(6)}(\mu)$, 
$L(\mu), T]$ is plotted for
the case of a 140 GeV Higgs mass $[T=2 \; ln (140/175.6)]$.  The four-loop
term within $S$ is evaluated through use of the Pad\'e estimate $c_0 =
1306$ (Table 1) in conjunction with eqs. (2.11-2.13) for the 
$RG$-accessible coefficients $c_1, c_2$, and $c_3$.  The running coupling
$\alpha_s (\mu)$ and running mass $m_t (\mu)$ occurring within (2.7) are
evolved via four-loop $\beta$ and $\gamma$-functions from physical
reference values $\alpha_s (M_z) = 0.119$ and $m_t (m_t) = 175.6 \; {\rm GeV}$
\cite{cea}.

Figure 1 shows that the four-loop expression eliminates virtually all of
the residual scale dependence still evident in the three-loop rate in
the region $M_t \geq \mu \geq 30 \; {\rm GeV}$. The four-loop rate is
observed to have a local minimum at $\mu = 43.5 \; {\rm GeV}$ and a weak local
maximum at $\mu = 89.5 \; {\rm GeV}$.  The values of $S[x(\mu), L(\mu), T]$ at
both of these points of minimal-sensitivity \cite{pms} differ by only
0.2\%, indicative of the flatness of $S$ between these two points.  In
Table 3, values for $S[x(\mu), L(\mu), T]$ as well as the term-by-term
series $=x^2[1+R_1 x + R_2 x^2 + R_3 x^3]$ within $S$ are displayed for a
variety of $\mu$-values of interest.  The table displays the relative size 
of successive terms $R_n x^n$ at the minimal-sensitivity points
$\mu=43.5$ and $89.5 \; {\rm GeV}$ in addition to the points $\mu=140 \; {\rm GeV}
(=M_H)$ and $\mu=175 \; {\rm GeV} (\cong M_t)$.  The relative magnitude of the
four-loop term $R_3 x^3$ is seen to be less than 1\% of the
leading term in the series (unity) over the entire region between
the minimal-sensitivity values of $\mu$.

Of particular interest in this range are those values of $\mu \; [47.0 \;
{\rm GeV}$ and $73.5 \; {\rm GeV}]$ at which the four-loop term $R_3 x^3$
effectively vanishes.  These values correspond to the two points in
Figure 1 at which the 3-loop and 4-loop curves cross.  The usual
approach towards extracting information from an asymptotic series
$\sum_{n=0} R_n x^n \; (R_0 = 1)$ is to sum only that series' decreasing
terms, {\it i.e.} to evaluate $\sum_{n=0}^{n'} R_n x^n$ 
by choosing $n'$ such that
$|R_{n'} x^{n'}|$ is a minimum.  By choosing $\mu$ so as to
have $R_3(\mu)$ {\it vanish}, one can then argue 
(for this choice of $\mu$) that $n' = 3$,
suggesting that such a value for $\mu$ is optimal for estimating
the series from its {\it known} terms.  Of course, such an interpretation
rests on the assumed increase of terms $|R_n x^n|$ subsequent to $n'=3$;
all we can really be certain of is that $|R_4 x^4|>|R_3 x^3| \; (=0)$ at such a
value of $\mu$. Nevertheless, Table 3 shows that both values of $\mu$
for which $R_3 \rightarrow 0$ lie between the two minimal-sensitivity
points $[\mu=43.5$ and $89.5 \; {\rm GeV}]$, and that the $\mu$-sensitive
factor $S[x(\mu), L(\mu), T]$ within the rate (2.6) varies by less than
0.2\% over this entire region.
Over the full range of $\mu$ values displayed in Table 3 
$(43.5 \; {\rm GeV} < \mu < 175 \; {\rm GeV})$, the four-loop series 
term $(R_3 x^3)$ varies between 0 and 5\% of the leading one-loop order term (unity). 
Surprisingly, however, the rate $S[x(\mu), L(\mu), T]$ displays a relative
spread of values $\Delta S/S \leq 0.4\%$ over the same range of $\mu$, indicative 
of a substantial reduction in residual scale dependence.

In Figures 2 and 3 we exhibit the residual scale dependence of $S[x(\mu)$,
$L(\mu),T]$ to three and four-loop order for Higgs masses of 100 GeV and 175 GeV, 
respectively. These figures show the same reduction in scale dependence evident in 
Figure 1 when $M_H = 140 \; {\rm GeV}$.  For the case
of $M_H = 100 \; {\rm GeV}$, for example, Table 4 shows that the 4-loop term
$R_3 x^3$ varies between zero and 10\% of the leading-order 1-loop term
between the local minimum at $\mu = 29 \; {\rm GeV}$ and $\mu \cong M_t$.
Over this same range of $\mu$, the four-loop values of $S[x(\mu),
L(\mu), T]$ remain within 1\% of each other.  For $M_H = 175 \; {\rm GeV}$, the
four-loop term is between zero and 3\% of the leading-order term
between the local minimum at $\mu = 48.5 \; {\rm GeV}$ and $\mu \cong M_t$,
whereas the full four-loop expression for $S[x(\mu), L(\mu), T]$
exhibits a relative spread of only 0.1\% [Table 5].

Such scale independence is, of course, a reflection of the 
$RG$-invariance (2.8) of $S[x(\mu), L(\mu), T]$, which has been 
utilized explicitly to obtain the coefficients
$c_{1-3}$.  Nevertheless, the coefficient $c_0$, which is {\it not}
perturbatively accessible via (2.8) but is obtainable (at present) only 
by Pad\'e approximant methods, appears to
be precisely what is required to eliminate virtually all residual 
$\mu$-dependence in the rate arising from truncation.  
For example, if $M_H = 100 \; {\rm GeV}$, the factor
$c_0 x^3 (\mu)$ in isolation contributes more than 10\% of the leading
order contribution (unity) for $\mu \stackrel{<}{_\sim} M_t$. Moreover, this 
contribution
{\it increases} as $\mu$ decreases.  Nevertheless, the particular choice
$c_0 = 2453$ appears to ensure that the overall spread in $S$ remains
within 1\%, despite the potentially large contribution to this spread
from $c_0 x^3 (\mu)$.  Evidently the $\mu$ dependence of this term serves to
cancel the residual $\mu$-dependence of the remaining terms in the
series. \footnote{The reduction of scale-dependence via Pad\'e approximant
methods is discussed in detail by Gardi \cite{ega}.}

This near cancellation of residual scale-dependence makes possible a set
of credible predictions for the rate (2.6).  In the fourth column of 
Table 6 we have
tabulated the $H \rightarrow gg$ decay rates for $M_H = 100, 125, 140, 150$, 
and $175 \; {\rm GeV}$ in the $m_b \rightarrow 0, \; M_H^2 < < 4M_t^2$ limit.  
The largest source of uncertainty for these predictions
is in the value for $\alpha_s (M_z) =$ ($0.119 \pm 0.002$ \cite{cea}), which should 
lead to 4\% uncertainty in the rates presented in Table 6. 

Fermion mass effects ({\it i.e.} the departures from the $m_b
\rightarrow 0, \; M_H^2 < < 4M_t^2$ assumptions implicit in the
derivation \cite{kgc} of the $H \rightarrow gg$ rate) can be
accommodated in leading order by replacing the factor of unity in (3.1)
with the following $m_b$- and $M_t$-sensitive terms \cite{sdg}:

\renewcommand{\theequation}{4.\arabic{equation}}
\setcounter{equation}{0}

\begin{equation}
1 \rightarrow 1+\delta_m = \frac{9}{16} \left[ (A_t+Re A_b)^2 + (Im
A_b)^2 \right],
\end{equation}
\begin{equation}
A_t = 2\left[ \tau_t + (\tau_t - 1) \left(sin^{-1}(\sqrt{\tau_t})
\right)^2 \right] / \tau_t^2, \; \; \; \tau_t \equiv M_H^2 / 4M_t^2,
\end{equation}
\begin{equation}
A_b = 2 \left[ \tau_b + (\tau_b - 1) f (\tau_b) \right] / \tau_b^2, \;
\; \; \tau_b \equiv M_H^2 / 4m_b^2,
\end{equation}
\begin{equation}
f(\tau) \equiv - \frac{1}{4} \left[ ln \left( \frac{1+\sqrt{1-1/\tau}}{1
- \sqrt{1-1/\tau}} \right) - i \pi\right]^2.
\end{equation}
The right hand side of (4.1) is easily seen to approach unity when $m_b
\rightarrow 0, \; \; M_H^2 / 4M_t^2 \rightarrow 0$:
\begin{equation}
\lim_{\tau_t \rightarrow 0} A_t = \frac{4}{3}; \; \; \lim_{\tau_b
\rightarrow \infty} A_b = 0.
\end{equation}
The leading mass correction $\delta_m$, as defined in (4.1), is tabulated 
for various Higgs boson masses (with PDG \cite{cea} fermion-mass values
$m_b = 4.2 \; {\rm GeV}$, $M_t = 175.6 \; {\rm GeV}$) in the fifth column of Table
6.  This lowest-order fermion mass correction can be incorporated into
the rate (2.6) by its inclusion into the series (3.1)
\begin{equation}
S = x^2 [1+R_1 x + R_2 x^2 + R_3 x^3 + \delta_m].
\end{equation}
The sixth column of Table 6 tabulates the $H \rightarrow gg$ rate with
this correction included.

Because $\delta_m$ is a leading order correction, there is genuine ambiguity 
as to whether "physical" or running masses should be incorporated into
this correction.  The former choice yields a manifestly scale-dependent
contribution $\Delta S = \delta_m^p x^2 (\mu)$ to the rate $S$.  One
could argue for the incorporation of {\it running masses} $m_b (\mu),
m_t (\mu)$ in (4.2) and (4.3); {\it i.e.}, $\Delta S = \delta_m (\mu)
x^2 (\mu)$, with $\delta_m (\mu)$ calculated via (4.1-4) with
$\tau_t = M_H^2 / 4m_t^2 (\mu)$ and $\tau_b = M_H^2 / 4m_b^2 (\mu)$, and
with $\mu$ identified consistently with the minimal-sensitivity scale
tabulated in Column 2 of Table 6.  In Column 7 of Table 6 we have
tabulated $\delta_m (\mu)$ utilizing 6-active-flavour running masses
referenced to $m_t (m_t) = 175.6 \; {\rm GeV}$, $m_b(m_b)=4.2 \; {\rm GeV}$, and in
Column 8 we have listed the corresponding $H \rightarrow gg$ decay
rates. Theoretical uncertainty associated with the masses utilized 
in the leading correction-factor $\delta_m$ is reflected in the differing 
rates of Column 6 and Column 8.  This discrepancy is seen to be at most a 
2\% effect. 

The leading order mass-correction factor $\delta_m$ in (4.6) is seen from a 
comparison of Table 6 to Tables 3-5 to be generally smaller than the three-loop 
term $R_2 x^2$ in (4.6), but somewhat larger than the four-loop term $R_3 x^3$ 
also appearing in (4.6) [$\delta_m$ {\it is} smaller than $R_3 x^3$ when $\mu=M_t$, as 
in Table 2]. Since the next order of fermion mass corrections is
suppressed by an additional power of $\alpha_s$, we anticipate such
next-to-leading-order fermion mass corrections to be well within 1\% of
the total rate.  In any case, both leading-order mass corrections
$\delta_m$ and the four-loop contribution $R_3 x^3$ we obtain are
sufficiently small for credible estimates of the $H \rightarrow gg$ rate
on the basis of the terms in (4.6).  It is also worth noting from Table
6 that $\delta_m^p$ itself is only 1\% of the leading-order one loop
term [normalized to unity in (4.6)] when $M_H = 140-150 \; {\rm GeV}$, and that
$\delta_m(\mu)$ is comparably small when $M_H = 150 \; {\rm GeV}$. Hence,
the rates tabulated in Table 6 are optimally valid for Higgs masses in
the 140-150 GeV range.

\section*{5.  The $H \rightarrow b \bar{b}$ rate}

\renewcommand{\theequation}{5.\arabic{equation}}
\setcounter{equation}{0}

The leading QCD scalar-current correlation function contributions to the 
$Higgs \rightarrow b \bar{b}$ decay rate are known from explicit
calculation to ${\cal{O}}(\alpha_s^3)$, with secondary ${\cal{O}}(m_b^2
/ M_H^2)$ power corrections known to ${\cal{O}}(\alpha_s^2)$, as given
in \cite{che}:
\begin{eqnarray}
\Gamma (H \rightarrow b \bar{b})& =& \frac{3G_F}{4\sqrt{2} \pi} M_H m_b^2
(M_H) \Biggl[ \Pi_{scalar} (\mu^2 = s = M_H^2) \Biggr. \nonumber \\
& &\qquad\qquad\qquad\qquad\qquad\qquad\Biggl. + \frac{m_b^2(M_H)}{M_H^2} \left( -6 - 40 \frac{\alpha_s
(M_H)}{\pi} - 87.72 \left( \frac{\alpha_s (M_H)}{\pi} \right)^2 \right)
\Biggr].
\end{eqnarray}
The factor $\Pi_{scalar}$ in (5.1) is the imaginary part of the QCD
correlation function for the quark-antiquark scalar current normalized
to unity in leading order \cite{che}:
\begin{eqnarray}
 \Pi_{scalar} [\mu, s, x(\mu)] 
 =  1& +& \left( \frac{17}{3} + 2 \; ln \mu^2/s \right) x(\mu)\nonumber\\ 
&+& \left( 29.1467 + \frac{263}{9} ln \left( \frac{\mu^2}{s} \right) +
\frac{47}{12} ln^2 \left( \frac{\mu^2}{s} \right) \right) x^2 (\mu)
\nonumber \\
& + & \left( 41.7576+238.381 \; ln \left( \frac{\mu^2}{s} \right) + 94.6759 \; ln^2
\left( \frac{\mu^2}{s} \right) + 7.61574 \; ln^3 \left( \frac{\mu^2}{s}
\right) \right)x^3(\mu) \nonumber \\
& + & \left( d_0+d_1 ln \left( \frac{\mu^2}{s} \right) + d_2 ln^2 \left(
\frac{\mu^2}{s} \right) + d_3 ln^3 \left( \frac{\mu^2}{s} \right) + d_4
ln^4 \left( \frac{\mu^2}{s} \right) \right) x^4 (\mu) \nonumber \\
& + & ...
\end{eqnarray}
where $x(\mu) = [\alpha_s(\mu)]_{n_f = 5} / \pi$, corresponding to five
active flavours.

In ref.[2], a detailed asymptotic Pad\'e-approximant procedure is
presented for estimating the coefficients $d_{0-4}$ in (5.2).  The
methodology is virtually identical to that of Section 3 above, except
that five moments are now calculated for $R_4[w]$, the coefficient of
$x^4(\mu)$, based upon the asymptotic error formula prediction \cite{esc}
\begin{equation}
R_4 = \frac{R_3^2[R_2^3 + R_1 R_2 R_3 - 2R_1^3 R_3]}{R_2[2R_2^3 - R_1^3
R_3 - R_1^2 R_2^2]}.
\end{equation}
The results, as tabulated in Table 3 of ref. [2], are
\begin{equation}
d_0 = 64.2, \; d_1 = 745, \; d_2 = 1180, \; d_3 = 253, \; d_4 = 15.4 .
\end{equation}
The factors $d_{1-4}$ may be obtained directly from RG-invariance of the
physical rate 
\begin{equation}
\mu^2 \frac{d}{d \mu^2} \left[ m_b^2 (\mu) \Pi_{scalar} \left[ \mu,
s,x(\mu) \right] \right] = 0,
\end{equation}
leading to the following values (also tabulated) in \cite{ces}):
\begin{equation}
d_1 = 791.52, \; d_2 = 1114.7, \; d_3 = 260.06, \; d_4 = 14.755 .
\end{equation}
The strong agreement between (5.6) and (5.4) suggests that the estimate
for $d_0$ in (5.4) is a credible one.  It is interesting to note that a
much more naive Pad\'e approach for estimating $d_0$, in which (5.3) was
applied directly to $\Pi_{scalar} (\mu^2 = s = M_H^2)$, yielded a value
for $d_0$ (=67.25 \cite{esc}) surprisingly consistent with the estimate
in (5.4).  Only the latter estimate exhibits sensitivity to the
logarithmic terms in $\Pi_{scalar}$, which all vanish when $\mu^2 = s$.

If we utilize the estimate for $d_0$ quoted in (5.4) within the $H
\rightarrow b \bar{b}$ rate, we find that the correlation-function
factor within (5.1) is given to ${\cal{O}}(\alpha_s^4)$ by
\begin{equation}
\Pi_{scalar} (\mu^2 = s = M_H^2) 
= 1 + \frac{17}{3}x (M_H) + 29.1467 \; x^2 (M_H) 
+ 41.7576 \; x^3 (M_H) + \underline{64} \; x^4 (M_H).
\end{equation}
The underlined coefficient is, of course, $d_0$ as estimated in
\cite{ces} via asymptotic Pad\'e-approximant methods.

There is genuine value in having an estimate of this term, as it is
generally larger than the ${\cal{O}}(\alpha_s^2 m_b^2 / M_H^2)$ final
term in (5.1).  For example, suppose that $M_H = 130 \; {\rm GeV}$, $m_b (130
\; {\rm GeV}) = 2.7 \; {\rm GeV}$ and $x(130 \; {\rm GeV}) = 0.114 / \pi$ (these values
are consistent with those employed in Table 1 of ref. \cite{che}).  The
relative magnitudes of the correlator contributions 95.5) to the $H
\rightarrow b \bar{b}$ rate are seen to be
\begin{equation}
\Pi_{scalar} \left( \mu^2 = s = (130 \; {\rm GeV})^2 \right) 
= 1 + 0.2056 + 0.0384 + 0.0020 + \underline{0.00011}.
\end{equation}
All but the underlined Pad\'e-estimate term are tabulated in Table 1 of
ref. \cite{che}.  Comparison to the power-suppressed terms in (5.1),
\begin{equation}
\frac{m_b^2(M_H)}{M_H^2} \left( - 6 - 40 \; x(M_H) - 87.72 \; x^2 (M_H)
\right) 
\begin{array}{c}{} \\ \longrightarrow \\_{M_H=130 \; {\rm GeV}}
\end{array}
- 0.0026 - 0.00062 - 0.00005,
\end{equation}
(also tabulated in \cite{che}) reveals that the ${\cal{O}}(\alpha_s^4)$
term in (5.8) is double the magnitude of the ${\cal{O}}(\alpha_s^2)$
term in (5.9). Hence the Pad\'e-estimated ${\cal{O}}(\alpha_s^4)$ term
in (5.7) enables one to utilize the full precision available in the
known power-suppressed contributions to (5.1).  However, one cannot
anticipate such precision experimentally for many years to come.

\section*{6.  Summary} 
        In the preceding sections, we have demonstrated how asymptotic
Pad\'e-approximant methods may be utilized to estimate the unknown
${\cal O}\left(\alpha_s^5\right)$ coefficients 
$\left\{c_0,~ c_1,~ c_2,~ c_3\right\}$ within the $H\rightarrow 2g$ decay
rate (2.7). Such estimates for $\left\{c_1,~ c_2,~ c_3\right\}$ are seen (Table 1) to be
within a few percent of the true values for these coefficients, which can
be extracted via renormalization-group methods.  This accuracy supports
corresponding asymptotic Pad\'e-approximant estimates of the
renormalization-group-inaccessible coefficients $c_0$ presented in Table 1.
Moreover, the inclusion of ${\cal O}\left(\alpha_s^5\right)$ terms within (2.7) is seen to
virtually eliminate the scale-parameter dependence of the rate over an
astonishingly large range of the scale-parameter $\mu$, typically 
$0.3 M_H  \stackrel{<}{_\sim} \mu  \stackrel{<}{_\sim}  M_t$. 
Inclusion of estimated ${\cal O}\left(\alpha_s^5\right)$ corrections in
conjunction with leading-order fermion-mass corrections to the rate (Table
6) are seen to reduce the perturbative uncertainty in the $H\rightarrow 2g$ decay rate
from ${\cal O}(20\%)$ to ${\cal O}(2\%)$.  Four-loop corrections to the 
$H \rightarrow b \bar b$ decay mode
are also presented, which are seen to reduce the very small perturbative
uncertainty of this dominant hadronic mode by an additional order of
magnitude.

\section*{Acknowledgement}

We are grateful for support from the Natural Sciences and Engineering
Research Council of Canada.

\newpage

\begin{table}
\begin{center}
\begin{tabular}{ccccc}
$M_H ({\rm GeV})$ & $\delta c_1 / c_1^{RG}$ & $\delta c_2 / c_2^{RG}$ & $\delta c_3 / c_3^{RG}$ & $c_0$\\
\hline
100 & -0.010 & -0.019 & -0.054 & 2453\\
125 & -0.009 & -0.013 & -0.059 & 1646\\
140 & -0.009 & -0.009 & -0.062 & 1306\\
150 & -0.009 & -0.007 & -0.068 & 1120\\
175 & -0.012 & -0.0004 & -0.070 & 763
\end{tabular}
\caption{The final column tabulates asymptotic Pad\'e-approximant predictions for the four loop nonlogarithmic
coefficient $c_0$ within (2.7), as obtained via moments of $c_0 - c_1 ln w + c_2 ln^2 w - c_3 ln^3 w$, for Higgs 
masses between 100 GeV and $M_t$.  Relative errors between predictions and true values of
$c_1, c_2$ and $c_3$, as determined from RG invariance, are tabulated in the second, third
 and fourth column:  $\delta c_i = \left(c_i^{predicted} - c_i^{RG}\right) / c_i^{RG}$.}
\end{center}
\end{table}

\begin{table}
\begin{center}
\begin{tabular}{c|c}
$M_H ({\rm GeV})$ &  $S[x(M_t), L(M_t),T]/x^2 =1+a_0 x + b_0 x^2 + c_0 x^3$\\
\hline
100 &  1 + 0.7728 + 0.3333 + \underline{0.1014}\\
125 &  1 + 0.7133 + 0.2675 + \underline{0.0679}\\
140 &  1 + 0.6831 + 0.2361 + \underline{0.0538}\\
150 &  1 + 0.6647 + 0.2177 + \underline{0.0461}\\
175 &  1 + 0.6237 + 0.1783 + \underline{0.0312}
\end{tabular}
\caption{The perturbative series $S/x^2$ within the $H \rightarrow 2g$ decay rate (2.7), 
evaluated at $\mu = m_t (\mu) = 175.6\,{\rm GeV}$.  $x(175.6 {\rm GeV}) = 0.10915/\pi$, consistent with
$x(M_z) = \alpha_s (M_z)/\pi = 0.119/\pi$. The final underlined term of the series
is obtained from the $c_0$ estimates in Table 1.}
\end{center}
\end{table}

\newpage

\begin{table}
\begin{center}
\begin{tabular}{c|c|c|c|c}
$\mu ({\rm GeV})$ & $\alpha_s(\mu)$ & $m_t(\mu)$$({\rm GeV})$ & $1+R_1x + R_2 x^2 + R_3 x^3$ & $S[x(\mu), L(\mu), T]$\\
\hline
 43.5 & 0.1325 & 198.2 & 1 + 0.3811 - 0.0395 + 0.0023 & 0.002390\\
 47.0 & 0.1309 & 196.8 & 1 + 0.4014 - 0.0251 - 0.0001 & 0.002390\\
 73.5 & 0.1226 & 188.8 & 1 + 0.5094 + 0.0614 + 0.0000 & 0.002393\\
 89.5 & 0.1193 & 185.6 & 1 + 0.5525 + 0.1007 + 0.0068 & 0.002394\\
  140 & 0.1124 & 178.8 & 1 + 0.6419 + 0.1907 + 0.0344 & 0.002390\\
  175 & 0.1092 & 175.6 & 1 + 0.6825 + 0.2354 + 0.0534 & 0.002384
\end{tabular}
\caption{The four-loop expression for the scale-dependent factor 
$S[x(\mu),L(\mu),T]$ for $M_H = 140 \; {\rm GeV}$ within the Higgs $\rightarrow$ gluon-gluon
decay rate (2.6) for five different values of $\mu$.  Weak extrema within the rate
(Fig. 1) occur at $\mu=43.5$ and $89.5 \; {\rm GeV}$.  Also displayed are $\mu=M_H$, 
$\mu \approx M_t$, and two values of $\mu$ ($47.0 \; {\rm GeV}$ and $73.5 \; {\rm GeV}$) at which
the four-loop contribution to $S$ is almost zero.  The values of $S$ over this entire 
range of $\mu$ are seen to be remarkably static, as discussed in the text.}
\end{center}
\end{table}

\begin{table}
\begin{center}
\begin{tabular}{c|c|c|c|c}
$\mu ({\rm GeV})$ &  $\alpha_s(\mu)$ & $m_t(\mu)$ $({\rm GeV})$ & $1+R_1x + R_2 x^2 + R_3 x^3$ & $S[x(\mu), L(\mu), T]$\\
\hline
 29.0 & 0.1413 & 206.5 & 1 + 0.3819 - 0.0546 + 0.0033 & 0.002691\\
 31.0 & 0.1397 & 205.1 & 1 + 0.4007 - 0.0410 + 0.0002 & 0.002691\\
 54.0 & 0.1282 & 194.2 & 1 + 0.5418 + 0.0775 + 0.0001 & 0.002698\\
 65.5 & 0.1247 & 190.8 & 1 + 0.5853 + 0.1198 + 0.0089 & 0.002699\\
  100 & 0.1175 & 183.8 & 1 + 0.6722 + 0.2125 + 0.0402 & 0.002693\\
  175 & 0.1092 & 175.6 & 1 + 0.7722 + 0.3326 + 0.1010 & 0.002667
\end{tabular}
\caption{The scale-dependence of the four-loop expression for $S[x(\mu),
L(\mu), T]$ is displayed, as in Table 3, for the case $M_H = 100 \;
{\rm GeV}$.  Weak extrema in Figure 2 occur at $\mu=29.0$ and $65.5 \; {\rm GeV}$.
The four-loop contribution is seen to nearly vanish at $\mu=31.0$
and $54.0 \; {\rm GeV}$.}
\end{center}
\end{table}

\newpage

\begin{table}
\begin{center}
\begin{tabular}{c|c|c|c|c}
$\mu ({\rm GeV})$ &  $\alpha_s(\mu)$ & $m_t(\mu)$ $({\rm GeV})$ & $1+R_1x + R_2 x^2 + R_3 x^3$ & $S[x(\mu), L(\mu), T]$\\
\hline
 48.5 & 0.1303 & 196.2 & 1 + 0.3385 - 0.0584 + 0.0081 & 0.002216\\
 61.0 & 0.1260 & 192.0 & 1 + 0.3976 - 0.0190 - 0.0001 & 0.002216\\
 91.0 & 0.1190 & 185.3 & 1 + 0.4912 + 0.0537 + 0.0000 & 0.002218\\
 111 & 0.1159 & 182.2 & 1 + 0.5338 + 0.0910 + 0.0058 & 0.002218\\
 175 & 0.1092 & 175.6 & 1 + 0.6230 + 0.1777 + 0.0309 & 0.002215
 \end{tabular}
\caption{The scale-dependence of the four-loop expression for $S[x(\mu),
L(\mu), T]$ is displayed, as in Table 3, for the case $M_H = 175 \;
{\rm GeV}$.  Weak extrema in Figure 3 occur at $\mu=48.5$ and $111.0 \; {\rm GeV}$.
The four-loop term is seen to nearly vanish at $\mu=61.0$
and $91.0 \; {\rm GeV}$. The relative spread in the rate $S$ is seen to be only
of order 0.1\% for values of $\mu$ between $48.5$ and $175 \; {\rm GeV}$.}
\end{center}
\end{table}

\begin{table}
\begin{center}
\begin{tabular}{cccccccc}
$M_H $ & $\mu$ & $S$ & $\Gamma_{H \rightarrow gg}^{\delta_m = 0}$ 
& $\delta_m^p$ & $\Gamma_{H \rightarrow gg}^{\delta_m^p}$ & $\delta_m (\mu)$ 
& $\Gamma_{H \rightarrow gg}^{\delta_m (\mu)}$ \\
\hline
100 & 65.5 & 0.00270 & $1.97 \cdot 10^{-4}$ & -0.085 & $1.87 \cdot 10^{-4}$ & -0.059 & $1.90 \cdot 10^{-4}$ \\
125 & 80.5 & 0.00249 & $3.55 \cdot 10^{-4}$ & -0.040 & $3.47 \cdot 10^{-4}$ & -0.015 & $3.52 \cdot 10^{-4}$\\
140 & 89.5 & 0.00239 & $4.79 \cdot 10^{-4}$ & -0.012 & $4.76 \cdot 10^{-4}$ &  +0.011 & $4.82 \cdot 10^{-4}$\\
150 & 95.5 & 0.00234 & $5.75 \cdot 10^{-4}$ & +0.0076 & $5.78 \cdot 10^{-4}$ & +0.023 & $5.83 \cdot 10^{-4}$\\
175 & 111 & 0.00222 & $8.67 \cdot 10^{-4}$ & +0.060 & $8.99 \cdot 10^{-4}$ &  +0.077 & $9.08 \cdot 10^{-4}$
\end{tabular}
\caption{The four-loop order $H \rightarrow gg$ decay rate (2.6) is tabulated
for various choices of $M_H$.  The second column lists the 
minimal-sensitivity values of $\mu \; (dS/d\mu = 0)$ closest to $M_H$.  The third
column is the value of $S[x(\mu), L(\mu), T]$ evaluated at this choice
for $\mu$, as indicated in the previous three tables.  The fourth column
is the rate in the $m_b \rightarrow 0, \; M_H^2 < < 4M_t^2$ limit.  
The fifth column tabulates the magnitudes of the leading fermion mass correction
to $S/x^2$, as discussed in the text, using "physical" (p) fermion masses 
$M_t = 175.6 \; {\rm GeV}$, $m_b = 4.2 \; {\rm GeV}$.The sixth column tabulates the
$H \rightarrow gg$ rate incorporating this mass correction. The seventh column
lists the leading fermion mass correction to $S/x^2$ utilizing running fermion mass values
$m_b (\mu), m_t (\mu)$ at the Column 2 values of $\mu$, and the final column tabulates the 
$H \rightarrow gg$ rate incorporating this fermion-mass correction.  All masses and decay
rates are in ${\rm GeV}$ units.}
\end{center}
\end{table}

\clearpage
\begin{figure}[hbt]
\centering
\includegraphics[scale=0.6, angle=270]{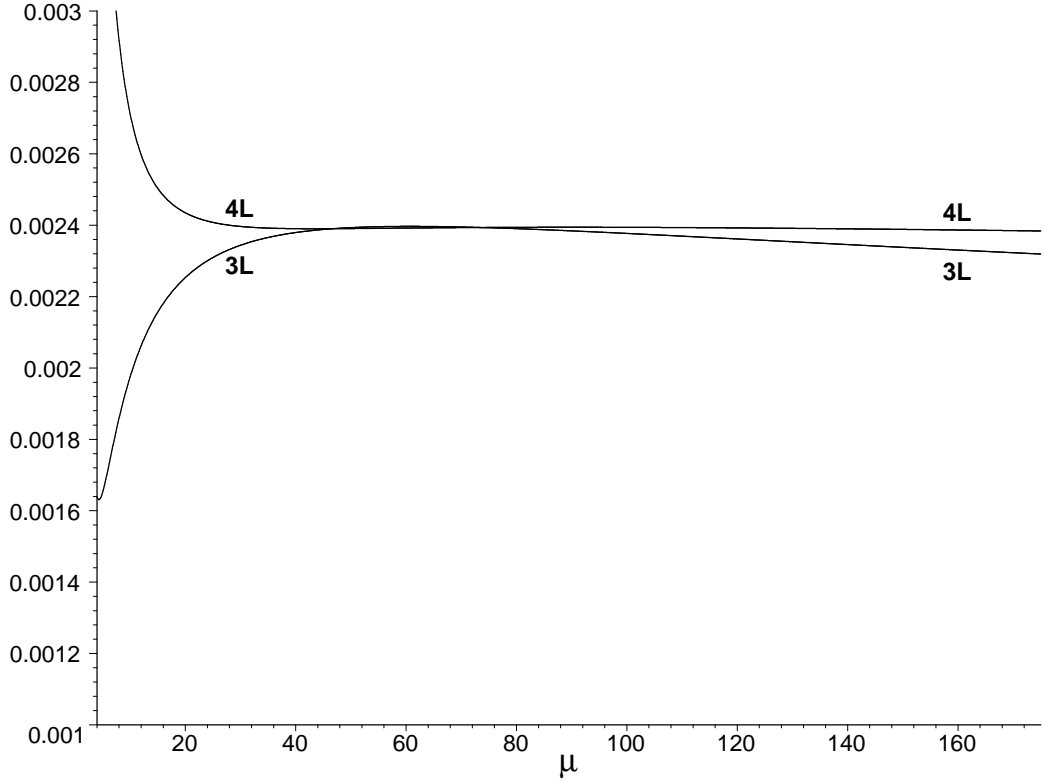}
\caption{
 Comparison of the residual renormalization scale- ($\mu-)$
dependence of the three-loop (3L) and four-loop (4L) expression for the
$H \rightarrow gg$ decay rate of a Higgs boson with mass $M_H = 140 \;
{\rm GeV} $.  
The $y$-axis numbers correspond to the dimensionless scale-dependent
factor $S\left[x\left(\mu\right),L\left(\mu\right), T\right]$ within the
rate (2.6), and the $x$-axis is the scale $\mu$ in GeV.
The logarithmic coefficients of 4L-contributions are
extracted via renormalization-group equation methods.  The 4L 
non-logarithmic coefficient is obtained via asymptotic Pad\'e approximant
methods.  The 3L and 4L curves nearly coincide in the region $44 \; {\rm GeV} 
\stackrel{<}{_\sim} \mu \stackrel{<}{_\sim} 80 \; {\rm GeV}$, indicating a
very small 4L contribution over this range of $\mu$.
}
\label{fig1}
\end{figure}

\clearpage
\begin{figure}[hbt]
\centering
\includegraphics[scale=0.6,angle=270]{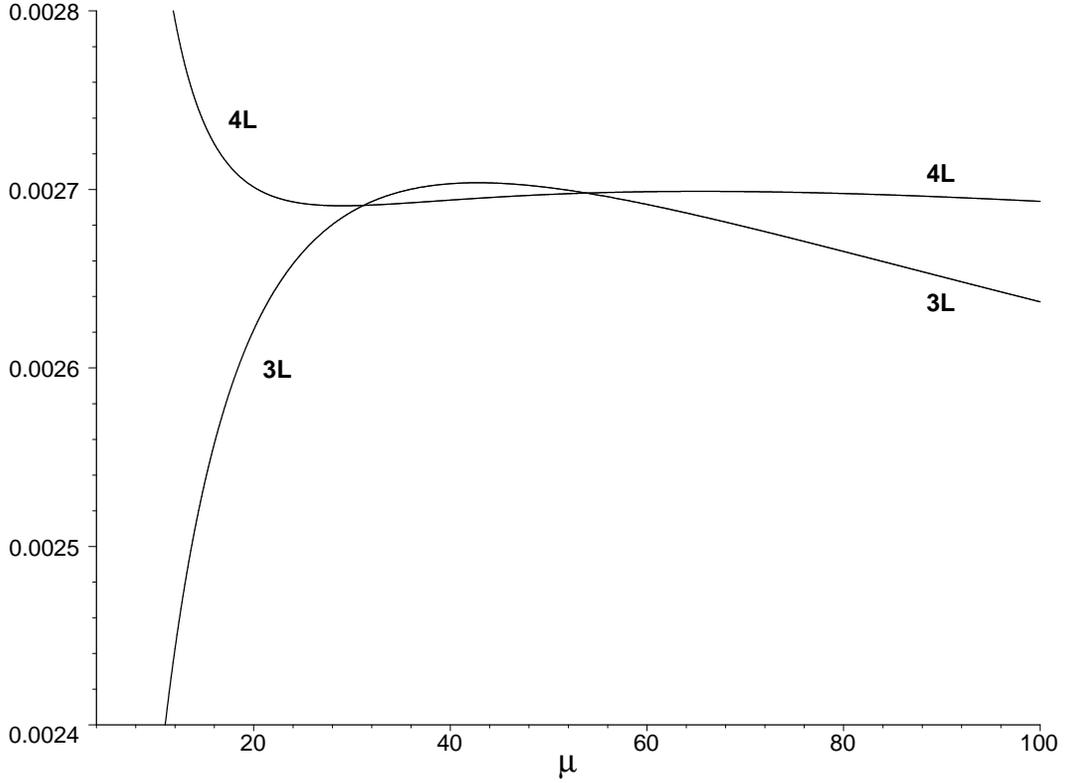}
\caption{
A comparison of the $\mu$-dependence of three-loop and
four-loop expressions for the $H \rightarrow gg$ rate is plotted, as in
Figure 1, but now for $M_H = 100 \; {\rm GeV} $.  
The $y$-axis numbers correspond to the dimensionless scale-dependent
factor $S\left[x\left(\mu\right),L\left(\mu\right), T\right]$ within the
rate (2.6), and the $x$-axis is the scale $\mu$ in GeV.
The curves cross at
values of $\mu$ for which the (estimated) four-loop contribution to the
rate is zero.
}
\label{fig2}
\end{figure}

\clearpage
\begin{figure}[hbt]
\centering
\includegraphics[scale=0.6,angle=270]{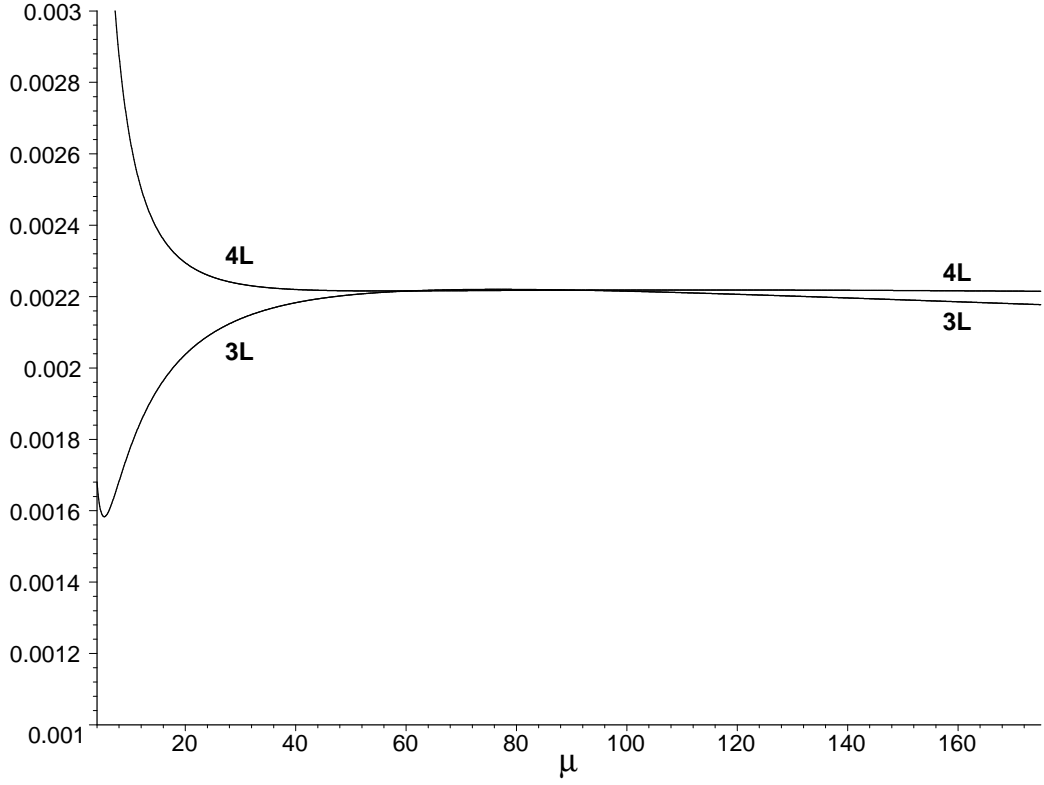}
\caption{
 A comparison of the $\mu$-dependence of three-loop and
four-loop expressions for the $H \rightarrow gg$ rate (as in Figs. 1 and
2), for $M_H = 175 \; {\rm GeV} $.
The $y$-axis numbers correspond to the dimensionless scale-dependent
factor $S\left[x\left(\mu\right),L\left(\mu\right), T\right]$ within the
rate (2.6), and the $x$-axis is the scale $\mu$ in GeV.
}
\label{fig3}
\end{figure}

\end{document}